# Changes in the Magnetization of a Double Quantum Dot.

T.H. Oosterkamp, S.F. Godijn, M.J. Uilenreef, Y.V. Nazarov,
N.C. van der Vaart and L.P. Kouwenhoven.
Department of Applied Physics and DIMES,
Delft University of Technology,
P.O.Box 5046,2600 GA Delft, The Netherlands

### Abstract

From accurate measurements of the energy states in a double quantum dot we deduce the change in magnetization due to single electron tunneling. As a function of magnetic field we observe crossings and anti-crossings in the energy spectrum. The change in magnetization exhibits wiggles as a function of magnetic field with maximum values of a few effective Bohr magnetons in GaAs. These wiggles are a measure of the chaotic motion of the discrete energy states versus magnetic field. Our results show good agreement with a numeric calculation but deviate significantly from semiclassical estimates.

Orbital magnetization of small electron systems has become an important issue in the field of mesoscopics, for instance in relation to the issue of persistent currents in rings [1]. Altshuler et al. [2] have pointed out that a non-zero orbital magnetization can be present in any mesoscopic electron system, regardless the precise geometry. The point of interest is that the magnetization measures the cumulative motion of the occupied quantum states as a function of magnetic field. Generally, this motion is chaotic, except for very specific conditions of separable geometries [5]. The statistical properties of the chaotic motion are supposed to be universal in the sense that they do not depend on the details of the microscopic structure. Direct measurements of the magnetization of a mesoscopic object is a challenging task, since it requires the detection of tiny magnetic moments [3].

In quantum dots the chaotic nature can be measured in electron transport [4]. For instance, fluctuations in the Coulomb peak heights have been measured and succesfully explained by random matrix theory (RMT) [6, 7]. Due to the non-interacting character of RMT, it has not been possible to describe the results of several studies on peak spacings (addition energies) [8]. Here, we report on an experimental study of the magnetization of a quantum dot. We show that semi-classical estimates can not explain our results, implying that a system with ∼50 electrons is too small to be described by RMT.

We measure the energy evolution versus $B$ of energy states near the Fermi energy $E_F$. The resolution is high enough that, for the first time, avoided crossings in the spectrum of a quantum dot can be resolved. We then obtain the magnetization by taking the derivative of energy with respect to $B$. Although this magnetization only includes contributions from states near $E_F$, this part largely determines the total magnetization [1]. Measurements of single-particle states versus $B$ have previously been reported on single quantum dot devices [9, 10, 4], but have not been analyzed in terms of their magnetization. In this paper we address a *double* quantum dot system which allows for a much better energy resolution compared to single dots. From the energy dependence on $B$ we calculate the magnetization. The advantage of our method is that the background magnetization of the whole heterostructure [12] is not measured so that we can concentrate on our mesoscopic system. As we explain below, we actually measure changes in the magnetization, $\Delta M$. We find that $\Delta M$ induced by one electron tunneling between the two dots is of order one effective Bohr magneton, $\mu_{GaAs} = e\hbar/2m_{GaAs} \simeq 0.87$ meV/T, which we determine with an accuracy of $0.1\,\mu_{GaAs}$. The magnitude of $\Delta M$ and the typical period of wiggles in $\Delta M$ as a function of $B$ are in good agreement with numerical calculations but, importantly, our results deviate from semiclassical estimates.

Figure 1a shows our double dot device. The metallic gates (1, 2, 3, and F) are fabricated on top of a GaAs/AlGaAs heterostructure with a 2DEG 100 nm below the surface. The 2DEG has a mobility of $2.3 \times 10^6$ cm$^2$/Vs and an electron density of $1.9 \times 10^{15}$ m$^{-2}$ at 4.2 K. From the density and the effective mass $m^*_{GaAs} = 0.067\,m_e$, follow the Fermi energy $E_F = 6.9$ meV and the Fermi wave vector $k_F = 1.1 \cdot 10^8$ m$^{-1}$. Applying negative voltages to all the gates depletes the electron gas underneath them and forms two weakly coupled quantum dots with an estimated size of 170 nm by 170 nm for the left dot and 130 nm by 130 nm for the right dot (lithographic sizes are (320 nm)$^2$ and (280 nm)$^2$). These dots contain about 60 and 35 electrons, respectively. The sample is cooled in a dilution refrigerator to 10 mK. Noise enhances the effective electron temperature in the 2D source and drain contacts to ∼80 mK. We measure the current in response to a dc voltage $V_{sd}$ applied between the source and drain contacts. The tunnel coupling between the dots and to the reservoirs can be controlled with the voltages on gates 1, 2 and 3. The experiments are performed in the weak coupling limit, meaning that mixing between quantum states in one dot with states in the other dot or in the leads is negligible.

In the weak coupling limit transport is governed by the physics of Coulomb blockade. We label the number of electrons in the left and right dot by ($N_l$,$N_r$). Tunneling between two dots occurs when certain conditions for the Coulomb energies are fulfilled and when *simultaneously* a quantum state in the left dot aligns with a state in the right dot [4]. We first discuss the conditions for the Coulomb energy. A transition from the left to the right dot can occur when the Coulomb energy of having $(N_l+1,N_r)$ exceeds the energy of $(N_l,N_r+1)$. To avoid transport through other charge states than $N_l$, $N_l+1$, $N_r$ and $N_r+1$, we choose a source-drain voltage which is just smaller than the smallest of the charging energies of the individual dots. The measured charging energies are $E_{C,left} = 1.2$ meV for the left dot and $E_{C,right} = 1.8$ meV for the right dot. We sweep the gate voltages over small ranges and focus on a particular charging transition; i.e. transitions between $(N_l+1,N_r)$ and $(N_l,N_r+1)$ only. Since we discuss only one transition at a time, we can, for simplicity, leave out the Coulomb energies from the discussion and concentrate on the alignment of quantum states.

Figure 1b illustrates the case where a quantum state of the left dot is aligned with a quantum state in the right dot; this is a case where current can flow. In contrast to resonant tunneling in a single dot, where the peak width is determined by the thermal broadening in the leads, the width of the current resonance in the double dot is determined only by the alignment of the quantum states. The measured resonance can be an order of magnitude narrower than the thermal energy $k_B T$ of the reservoirs [11]. We use this advantage of high energy resolution in a double dot system to obtain the magnetization with very high precision.

The quantum states (dotted and solid lines in Fig. 1b) we deal with are real many-body states of the dot systems. General labels for these states are $E_i^{l,N_l}$ for the left dot and $E_i^{r,N_r}$ for the right dot, which we simplify to $E_i^l$ and $E_i^r$. When sweeping gate 3 the condition for tunneling between the lowest possible states, i.e. from ground state to ground state, is $E_0^l = E_0^r - \alpha V_{g3}$, where $\alpha$ describes the influence of gate 3 on the right level. The situation of Fig. 1b corresponds to tunneling from the first excited state to ground state, i.e. $E_1^l = E_0^r - \alpha V_{g3}$. The states $E_i^l$ and $E_i^r$, i = 0, 1,2,$\cdots$ are separated by $\sim$ 150-200 $\mu$eV.

Figure 1c shows a typical set of current traces for different magnetic fields while sweeping the voltage on gate 3. The bias voltage $V_{SD} = 1.2$ mV such that several energy states in each dot are between the Fermi energies of the two leads. The *change* in peak positions versus $B$ is proportional to the motion of the alignment $\delta[E_i^l - E_i^r] = -\alpha \delta V_g^{peak}$. (Note that if the states $E_i^l$ and $E_i^r$ have the same $B$-dependence, the peak would not change its position.) We determined the factor $\alpha = 63\,\mu\text{eV}/\text{mV}_g$ through independent measurements from which we deduced the energy scale for Fig. 1c that is indicated by the arrow in the lower left inset ($\alpha$ does not change in this magnetic field range). The energy resolution of $[E_i^l - E_i^r]$ is $\sim$5 $\mu$eV.

The data in Fig. 1c contains several interesting features. First, we observe crossings between different peaks as well as anti-crossings (two are indicated by arrows). Second, pairs of peaks exhibit the same $B-$dependence. These are general features that we observe at many charge transitions ($N_l$,$N_r$). We do not know of similar observations in other quantum dot experiments. Independent measurements on *one* of the individual dots also showed states evolving in pairs below $B \sim 0.5$ T. The observed pairing and (anti)crossing of the peaks in the double dot experiments can then be explained as shown schematically in Fig. 1d. Suppose two energy states in one dot have an anti-crossing in their $B$-dependence. Then two paired energy states in the other dot, having the same $B$-dependence, will both probe this anti-crossing. At the points where two peaks actually cross two states in the left dot align with two states in the right dot simultaneously (though only one electron can tunnel at a time due to Coulomb blockade). These considerations explain our observations. The energy difference between paired states can be an exchange energy; e.g. when the higher energy state has spin zero and the lower state has spin one. In our experiments we find energy separations between paired states of typically 100 $\mu$eV being constant within 20 $\mu$eV over a field range of 0.5 T. Whether an exchange energy is giving rise to the energy separation is yet unclear. However, there have been other indications that exchange-correlation plays an important role in quantum dots [10, 13].

In Fig. 2a measurements of the current as the magnetic field is increased up to 750 mT are shown in a grayscale representation for different electron numbers ($N_l$,$N_r$). The peak heights tend to vanish with higher magnetic fields, which limits the magnetic field range of our experiments. In Fig. 2b the peak positions are extracted and the gate voltage axis is converted to energy. Note that this data also shows crossings and anti-crossings. We obtain the change in magnetization, $\Delta M = M_{N_l+1,N_r,E_i^l} - M_{N_l,N_r+1,E_{i'}^r}$, by numerical differentiation:

$$\Delta M = -\frac{\delta[E_i^l - E_{i'}^r]}{\delta B} = -\alpha \frac{\delta V_g^{peak}}{\delta B}. \tag{1}$$

Figure 2c shows examples of $\Delta M$ in units of $\mu_{GaAs}$ versus $B$, for several alignments of different discrete energy states. $\Delta M$ shows wiggles with an amplitude of the order of $\mu_{GaAs}$ and a typical period of 0.3 T to 0.4 T. Note that a magnetic field $B_f$ between 0.1 and 0.2 T corresponds to one flux quantum penetrating the area of a single dot. Measurements such as those in Figures 1 and 2 were performed in three separate cooldowns and also for different electron numbers in the two dots. The magnetization at zero magnetic field is always zero due to symmetry $E_i(B) = E_i(-B)$. At finite magnetic fields, the sign of $\Delta M$ can be positive as well as negative.

We can estimate the magnitude and the period of magnetization oscillations from semi-classical theory. The amplitude $\Delta M$ is roughly the magnetic moment of one electron moving through a dot of size $a$ with momentum $p_F$, $M_{sc} = I_{sc} \cdot S \simeq \mu_{GaAs} p_F a/2\hbar$, where $I_{sc}$ is the current due to one electron encircling an area $S$ in the dot. In our case, $p_F a/2\hbar \simeq 9$, and thus $\Delta M \simeq 9\,\mu_{GaAs}$, which is several times larger than what we observe. The characteristic period of the wiggles $\Delta B$ can be estimated by equating magnetic energy $\Delta M \Delta B$ with the mean spacing between particle-in-a-box states, $\Delta E = \pi^2 \hbar^2 / 2 m_{GaAs} a^2$. This yields $\Delta B \simeq B_f \pi \hbar / p_F a \simeq 0.02 T$, which is an order of magnitude smaller than seen in our experiment.

To comprehend this discrepancy we have performed numerical simulations. The results are plotted as solid lines in Fig. 3 for three different pairs of levels. We regard the dots as squares with sides of 170 and 130 nm. To lift degeneracies characteristic for the square geometry and to account for probable disorder, we add a custom random potential to the dot potential. The random potential is formed by several rectangular wells of random size and position and with a typical depth of $\sim 0.1\,E_F$. Solving the Schrödinger equation we have calculated the energy levels in both dots versus $B$. From the difference of the magnetization of states close to the Fermi level in each of the two dots we obtain $\Delta M$. We have checked that the wiggles of the magnetization change are random depending on the realization of the potential. They do retain the same order of magnitude and the same typical period. To illustrate that the numerical results show a better agreement with the experimental data than the semiclassical estimate as far as the amplitude and the wiggle period is concerned, we have plotted magnetization measurements from Fig. 1 (squares and diamonds) and yet another data set (circles) that happen to fit the calculations reasonably well, even though, due to the random nature of the disorder, an exact fit is not expected.

For larger electron numbers ($N > 200$) our simulations begin to show agreement with the semi-classical estimates. This drives us to the conclusion that quantum dots with $N < 100$ are too small to be satisfactorily described by semi-classical theory.

In conclusion, we used a new method to explore magnetic properties of an ultra small system by means of an accurate transport measurement. The high resolution in energy made it possible to observe clear avoided crossings of states. For the magnetization an accuracy of $\sim 0.1\ \mu_{GaAs}$ was achieved. Magnetization traces manifest the chaotic motion of energy levels in magnetic field. The system appears to be too small for this motion to be described by semi-classical theory. There is, however, good agreement with the results of numerical simulations that incorporate the microscopic description of the dot. We believe that this implies that semi-classical random matrix theories can not be applied to describe the magnetization (or fluctuations in peak spacings) of quantum dots containing fewer than 100 electrons.

We thank Philips Laboratories and C. T. Foxon for providing the heterostructures and S. Cronenwett, C. Harmans, J. E. Mooij, C. W. J. Beenakker, M. Büttiker, G. Blatter, and A. I. Larkin for very instructive discussions. The work was supported by the Dutch Foundation for Fundamental Research on Matter

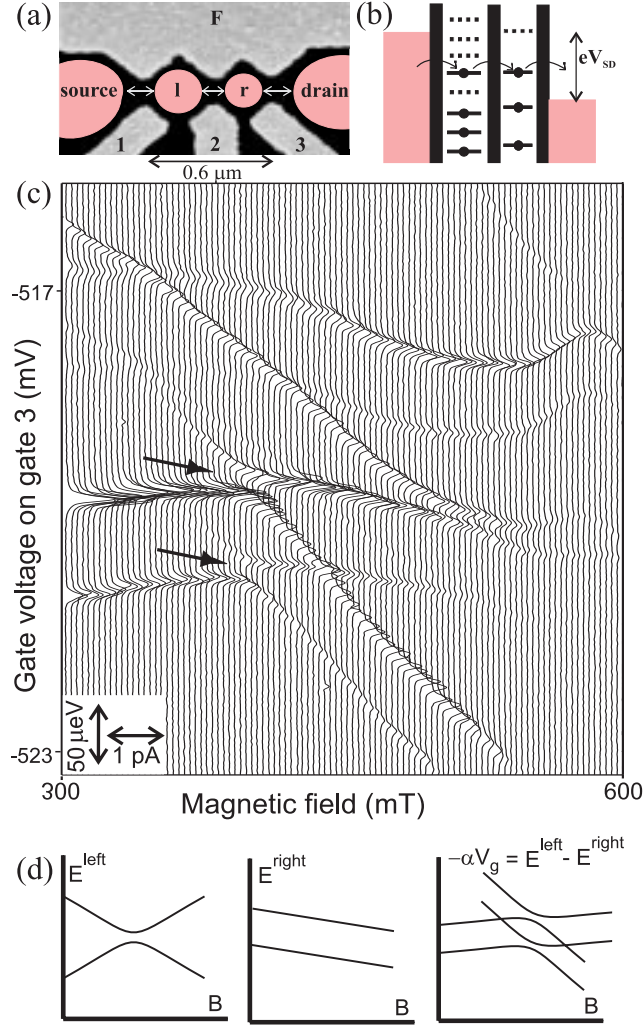

Figure 1: a) Geometrical layout of a double quantum dot: The gates are labeled F, 1, 2, and 3. Areas that form dots and bulk 2DEG are indicated by source, l, r, and drain. b) The energy diagram, uses black circles to indicate that a certain one-electron level is filled. Energy states at matching levels indicate that the electron can be transferred between the dots, which gives a peak in the current. Arrows illustrate the subsequent electron transfer through the system. c) Results of a typical measurement. We measure the current through the double dot sweeping the gate voltage at different magnetic fields. The curves are offset for clarity. From the leftmost to the rightmost curve, the magnetic field increases from 300 mT to 600 mT in 3 mT increments. d) The first two diagrams show how levels may evolve in each of the two dots as a function of magnetic field. When these four levels are scanned along each other by sweeping the gate voltage this will result in peak positions as sketched in the rightmost diagram.

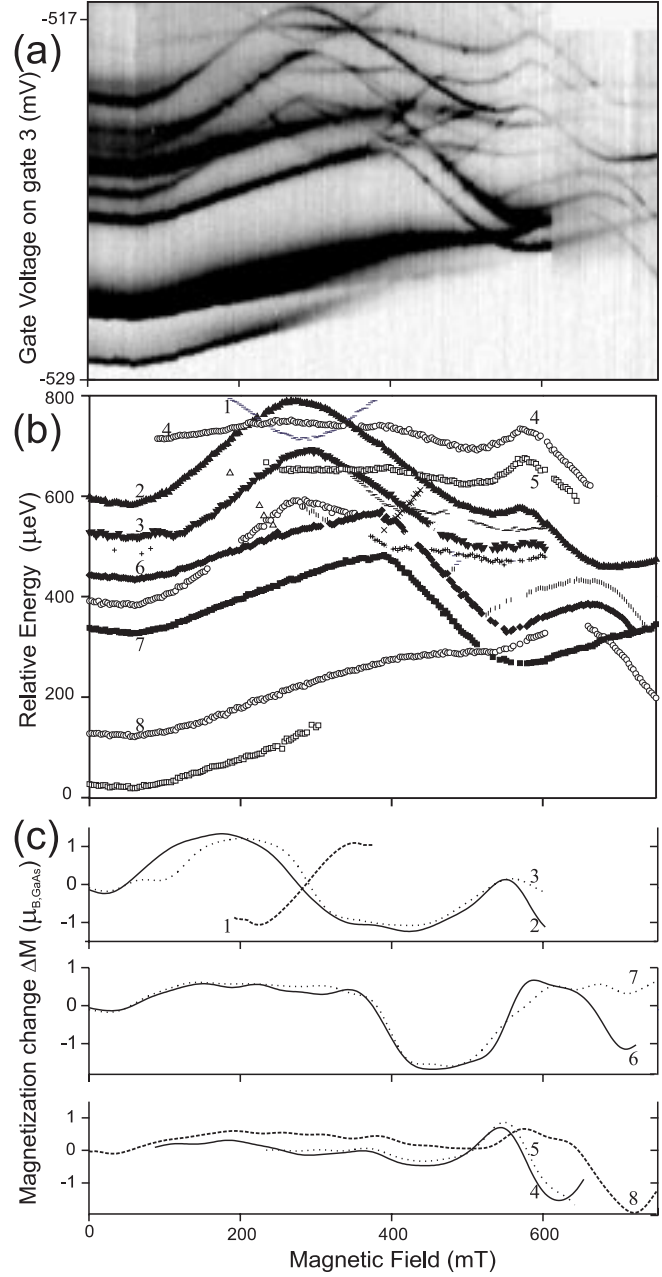

Figure 2: a) Grayscale representation of the current as a function of magnetic field. White is 0 fA, black is $\geq 150$ fA. Peaks larger than 150 fA appear broader in the grey scale plot because they are truncated. b) Peakpositions converted to energy. c) Magnetization change $\Delta M$ in units of $\mu_{GaAs}$ as calculated from the numbered curves in b) . Solid and dotted lines in one graph are taken from pairs: (2,3), (6,7) and (4,5).

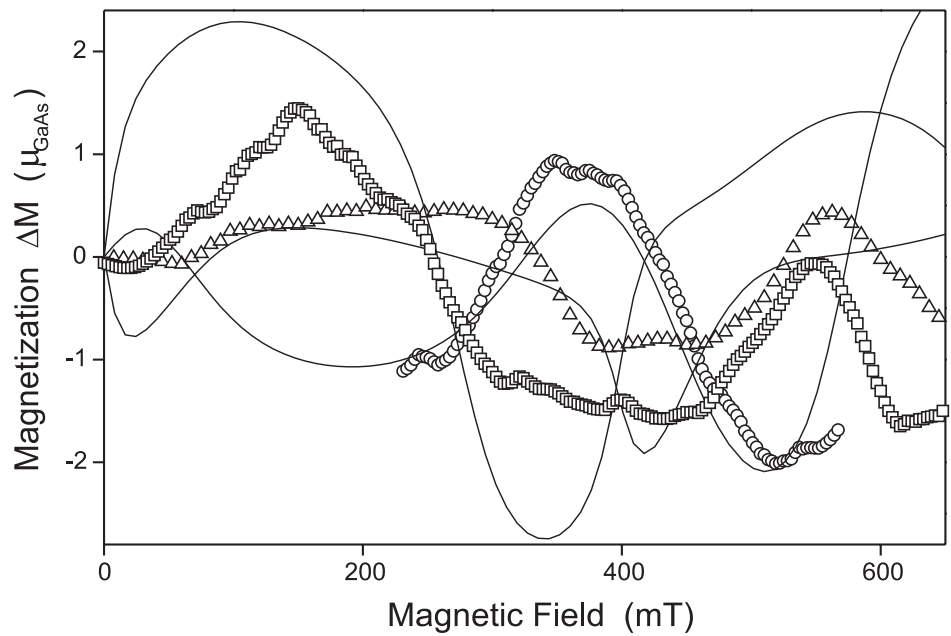

Figure 3: Magnetization change calculated numerically for two dots of the same sizes as in the experiment (solid lines) and experimental curves extracted from Fig. 1 (squares and triangles) and another dataset (circles).